\documentclass[12pt]{article}

\usepackage{graphicx}
\usepackage{amssymb}
\usepackage{epstopdf}
\usepackage{amsfonts}
\usepackage{amsmath}
\usepackage{bm}
\usepackage{mathrsfs}
\usepackage{subfigure}

\def\a{\alpha}
\def\b{\beta}

\def\ha{\frac{1}{2}}
\def\be{\begin{equation}}
\def\te{\end{equation}}
\def\ee{\end{equation}}
\def\nn{\nonumber}
\def\bea{\begin{eqnarray}}
\def\tea{\end{eqnarray}}
\def\eea{\end{eqnarray}}

\begin{document}


\title{Weyl Curvature Hypothesis in light of Quantum Backreaction at Cosmological Singularities or Bounces}
\author{
Bei-Lok Hu\\
{\small {\it Maryland Center for Fundamental Physics and Joint Quantum Institute}} \\
{\small {\it University of Maryland, College Park, Maryland 20742-4111, USA} [blhu@umd.edu]}
}
\date{\small (Oct. 28, 2021. v.2)}
 
\maketitle

-- In celebration of the 90th birthday of Professor Roger Penrose.  \\

-- Invited Paper in  {\it Universe: Special issue on Quantum Cosmology}, editor: Paulo Vargas Moniz.  
  


\newpage
\begin{abstract}
The Weyl curvature constitutes the radiative sector of the Riemann curvature tensor and gives a measure of the anisotropy and inhomogeneities of spacetime.  Penrose's 1979 Weyl curvature hypothesis (WCH) \cite{WCH}  assumes that the universe began at a very low gravitational entropy state,  corresponding to  zero Weyl curvature, namely,  the Friedmann--Lema\^{\i}tre--Robertson--Walker (FLRW) universe.  This is a simple assumption with far-reaching implications.  In classical general relativity,  Belinsky, Khalatnikov and Lifshitz (BKL) \cite{BKL} showed in the 70s that the most general cosmological solutions of the Einstein equation are that of the inhomogeneous Kasner types,  with intermittent alteration of the one direction of contraction (in the cosmological expansion phase), according to the  mixmaster dynamics of Misner (M) \cite{mix}. How could WCH and BKL-M co-exist? An answer was provided in the 80s with the consideration of quantum field processes such as vacuum particle creation \cite{Par69,Zel70}, which was copious at the Planck time ($10^{-43} sec$),  and their backreaction effects were shown to be so powerful as to rapidly damp away the irregularities in the geometry. It was proposed that  the vaccum viscosity due to particle creation \cite{HuPLA82} can act  as an efficient transducer of gravitational entropy \cite{HuPLA83} (large for BKL-M) to matter entropy,   keeping the universe at that very early time in a state commensurate with the WCH.
In this essay I expand the scope of that inquiry to a broader range, asking how  the WCH would fare  with various cosmological theories, from classical to semiclassical to quantum, focusing on their predictions near the cosmological singularities (past and future) or avoidance thereof, allowing the Universe to encounter different scenarios, such as undergoing a phase transition or a bounce.  WCH is of special importance to cyclic cosmologies, because any slight irregularity toward the end of one cycle will generate greater anisotropy and inhomogeneities in the next cycle.  We point out that regardless of what other processes may be present near the beginning and the end states of the universe, the backreaction effects of quantum field  processes  probably serve as the best guarantor of WCH because these vacuum  processes are ubiquitous, powerful and efficient in dissipating the irregularities to effectively nudge the Universe to a near-zero Weyl curvature condition. 
\end{abstract}
\newpage
\tableofcontents

\newpage

\noindent {\bf Preface} -- {\small When Professor Moniz  invited me last year to write a paper for this  special issue with some review or perspective emphasis  I could not identify a topic of current interest and importance in quantum cosmology on which I have done enough work to write about.   Then in July, Professor Hendrik Ulbricht brought to my attention that this August  is Professor  Penrose's 90th Birthday (b. August 8, 1931).  That got me thinking through some important themes in cosmology  from Penrose's viewpoint -- his concerns in a broad range of research topics in gravitation, quantum physics, black holes and cosmology, as lucidly presented in many semi-popular books (the last two  being \cite{CoT,FFF}),  are all of fundamental interest.  This is how I closed in to the title theme of this paper: The Weyl Curvature Hypothesis (WCH) \cite{WCH,Penrose2006,Penrose2008,Penrose2018}.}

{\small How does WCH fare when quantum backreaction is added to our considerations?   Backreaction in gravitational physics enters at four levels: classical, semiclassical, stochastic and quantum gravity.   In classical gravity it refers to the effects of the inhomogeneous modes on the homogeneous or long  wavelength (infrared) modes;  in quantum cosmology it refers to the effects of the inhomogeneous modes truncated in a mini-superspace approximation.     Backreaction at the semiclassical and stochastic levels has a different meaning.  It refers to the effects of quantum field processes on the structure and dynamics of a classical background spacetime.  This is a subject which I explored between 1977-1983-1995 in my work first with Parker \cite{HuPar77,HuPar78} and Hartle \cite{FHH79,HarHu79} from 77-80 (based on \cite{Hartle77},   continued by Hartle \cite{Hartle80,Hartle81} and Anderson \cite{And83,And84} from 1980-84),  then upgraded to a  closed-time-path,  in-in, or Schwinger-Keldysh treatment with Calzetta in 1987 \cite{CalHu87},  including noise and fluctuations in 1994 \cite{CalHu94},  with Matacz  in 1994 \cite{HuMatELE} and with Sinha in 1995 \cite{HuSinFDR}. This was continued by Campos and Verdaguer in 1994-1996 \cite{CamVer94,CamVer96}. These works explored the backreaction effects of the trace anomaly and particle creation in the early universe near the Planck time,  at the threshold of quantum gravity (defined as theories for the microscopic constituents of spacetime and matter).  The Weyl curvature hypothesis  (WCH)  \cite{WCH} will be examined in the context of semiclassical and stochastic gravity \cite{HuVer20} in Sec. 5.}

{\small Taking the time to write this essay has been a good way to  update my knowledge since my  1982-83 papers  on vacuum viscosity \cite{HuPLA82} and gravitational entropy \cite{HuPLA83} in the areas of  gravitational entropy, the Weyl curvature hypothesis, the cosmological singularity or avoidance of singularity, as predicted by different theories, and how vacuum viscosity due to the backreaction of  quantum field processes at the Planck time, such as pertaining to the trace anomaly and vacuum particle creation, would bear on these issues.  I wanted to see how the WCH  fares with other later developed models such as the cyclic cosmology of classical general relativity (GR) (Sec. 2.1) or particle physics (Sec. 2.2) origins,  singularity avoidance solutions from quantum cosmology (Sec. 3.1, 3.2),  in particular,  the `Big Bounce' solutions of loop quantum cosmology (Sec. 3.3) and quantum phase transitions in asymptotic freedom and causal dynamical triangulation (Sec. 4.1) -- whether there are  indications that the WCH holds in these contexts and, in some cases, whether it has proven to be desirable or even necessary.  (If the reader only wants to find out how the WCH and gravitational entropy fare in semiclassical gravity,  he/she can just selectively read Sec. 1.2, 2.3, 3.1 and 5.)}

{\small It is not my intention to cover the full range of topics mentioned above, aided with extensive literature, like a review.  Far from it.  I won't even repeat what Penrose said in his original proposal of the WCH and how he defined gravitational entropy,  or how he  argued for a conformal 
cyclic cosmology  in classical general relativity.  Instead I would urge  interested readers to read his original papers for insights which can never be reproduced.  For papers by other authors which I find relevant to our  central themes here  I also prefer to keep their original words rather than add my paraphrase or re-statement.  
The nature of this article is probably closer to  an outline, a selective highlight, a personal guide, as the topics chosen certainly reflect my own interests or what I view as important, which could be very different from other practitioners'.  }

\section{Weyl Curvature Hypothesis and Gravitational Entropy }

Penrose pointed out in his 1979 essay \cite{WCH} that if our Universe had begun close to what the Friedmann--Lema\^{\i}tre--Robertson--Walker ({FLRW})  models describe  -- in a state of highest degree of isotropy and homogeneity, corresponding to zero Weyl curvature -- it is something extremely special.  He reached this conclusion by comparing the entropy of matter we see today,  measured by the Hawking radiation, emitted by black holes, to the entropy of gravity, measured by the Weyl curvature tensor squared integrated over 3-volumes.  Gravitational entropy (GEnt) defined as such would increase in time from zero in the beginning (which could be a singularity, the `Big Bang',  or a state of minimal volume, a `bounce', see below) and provide a cosmological arrow of time.   
Penrose’s proposals for the non-activation of gravitational degrees of freedom registered in the Weyl curvature,  at the Big Bang,  are referred to as the Weyl Curvature Hypothesis (WCH).

Penrose's  position  on this theme has not changed in 40 years, as the following statement from his 2018 essay shows \cite{Penrose2018}   “In such a situation (the universe in a collapsing phase), we expect density irregularities to increase as the universe contracts, and the FLRW approximation would get worse and worse. In the late stages of the collapse, these irregularities would result in numerous black holes. Time reversing this picture (and taking into account the invariance of Einstein’s equations under time-reversal), we find that the singular origin that we find in the FLRW models is something extremely special."
While today we see the Planck distribution of matter suggesting a state of  maximum matter entropy,  “In contrast with the behaviour of a gas in a box, for example, where maximum entropy would be pictured as something with great spatial uniformity, gravitating bodies, such as systems of stars, would tend to clump more and more in their spatial distribution, as their dynamical time evolution proceeds, representing an increase in the gravitational entropy."

The four themes I listed above are already laden in these descriptions:  WCH, GEnt,  also the Bang-vs-Bounce issue.  The reason is, for people who prefer to see the universe not going to a singularity,  they have to come up with some suitable ways to avoid it, such as through processes in particle physics or in quantum gravity. We mention four of them in this essay:  pre-Big Bang,  loop quantum cosmology (LQC), causal dynamical triangulation (CDT)  and asymptotic freedom (AsyF)\footnote{This list is far from exhaustive. We are not treating inflationary cosmology \cite{Guth,InfCos} here since it already has a wide coverage and how the WCH fits in with  inflation is rarely discussed. On the relation of inflation and singularity,  Borde and Vilenkin \cite{BorVil} made a categorical statement that a physically reasonable spacetime that is eternally inflating to the future must possess an initial singularity. Vilenkin \cite{VilEtInfQCos} also questioned the necessity of quantum cosmology in the face of eternal inflation.  As for the speculations into our universe's future, there is considerable amount of work based on two types of inflationary models, one is the `stochastic inflation' of Starobinsky \cite{StoInf,StarFuture} which allows the noise associated with the fluctuations of  a quantum field  to drive the universe to inflation.  The other is the `eternal inflation' proposed and developed by Linde, Vilenkin, Guth and others \cite{EtInf}, where a universe can live forever with  the continual births of baby universes in the sequential and parallel generation of multiple branches.   For a recent assessment of the likelihood for eternal inflation in a variety of popular models, including the swampland of string theory,  see, e.g., \cite{NoEtInf}.}

For people who want  our future universe not just to bounce once or twice, but to continue  to bounce  forever without suffering much attenuation,  or stumbling  into a singularity, in the so-called `cyclic cosmologies',  the requirements are even more  stringent. What happens to the universe at the end of the contraction phase requires extra careful scrutiny.  The highly regular FLRW type of  approach to the singularity or bounce  would be preferred, because only then would the deviations be kept under control cycle after cycle.  This is in turn because gravitational forces tend to  clump irregularities,  so the next cycle would be more difficult to control than the last.  

Most people would agree that quantum effects are important at the beginning and ending stages of the universe,  either entering a singularity or with a minimal scale.  These developments and considerations add new impetus and interests to Penrose's WCH-GEnt, originally formulated without quantum considerations.  To the above mentioned subjects we shall also add  semiclassical and   quantum backreaction considerations. The former refers to the backreaction of quantum field processes on the dynamics of spacetime.  They need be included because these processes were dominant and had exerted significant effects at the Planck scale. The latter  pertains to LQC:  It concerns the effects of the inhomogeneous quantum modes in the superspace of 3-geometries,  so far  largely ignored in the mini-superspace formulations of  LQC.  

\subsection{Classical GR}  
Before we begin our studies of WCH in a quantum setting it is important to remind ourselves of the two very important theoretical achievements in the 60s in classical general relativity (GR), which underlie all investigations in black holes and cosmology built upon the GR theoretical framework:  the singularity theorems from 1965-68 of Penrose \cite{Penrose65}  , Hawking \cite{Hawking66} and Geroch \cite{Geroch66}, 
and the general solutions of the Einstein equations near the cosmological singularity in the work of Belinsky,  Khalatnikov,  Lifshitz (BLK) from the early 60s  (when the authors preferred a singularity-free scenario, later corrected, after the Penrose-Hawking singularity theorems) to the early 1970s \cite{BKL}   with the now famous BKL approach to cosmological singularity,  and  Misner's mixmaster universe \cite{mix} and minisuperspace quantum cosmology \cite{mss} from 1969 to 1972,  and the `velocity-dominated' approach by Eardley,  Liang and Sachs \cite{ELSachs}.

\subsubsection{Singularity Theorems: Penrose-Hawking-Ellis-Geroch}

The Penrose 1965 singularity theorem for black holes extended by Hawking to cosmological singularities are known as the Penrose-Hawking singularity theorems in general relativity \cite{HawPen}. They are  described in more pedagogical terms in the book by Hawking and Ellis \cite{HawEll}, the premium monograph on global analysis methods in general relativity. These theorems preclude  singularity-free “bounce” from occurring in the presence of any type of matter source with stress-energy tensor satisfying a very general condition of local energy non-negativity.  For a detailed  explanation of its contents with an extensive bibliography on this fundamental topic, see the 2015 review of \cite{SenGar15} in a mid-century celebration of Penrose's 1965 paper.

\subsubsection{Approach to Cosmological Singularity: BKL-Misner} 

In addition to the primary sources of the Belinsky-Khalatnikov-Lifshitz -Misner enterprise mentioned above,  
the 1975 book by Ryan and Shepley \cite{RyaShe} has an excellent introduction to spatially-homogeneous (Bianchi) cosmologies. See also the review of Barrow and Tipler \cite{BarTip}. For a discussion of the classical   BKL-M mixmaster chaotic dynamics (billiard ball in minisuperspace), see, e.g.,\cite{Barrow,Cronish}\footnote{Chaotic dynamics of minisuperspace cosmology is not limited to the mixmaster type.  As  Calzetta and Hasi show \cite{CalHas} it can also occur in  the dynamics of a spatially-closed FLRW universe conformally coupled to a real, free, massive scalar field  for large enough field amplitudes. }    For the latest developments of this respect  in relation to indefinite Kac-Moody algebra, see \cite{BelHen}.  
The role of chaos in the decoherence of quantum cosmology and the appearance of a classical spacetime has been pointed out by Calzetta \cite{CalChaosQCDec}.

Since many topics we shall discuss are related to the state of the universe near the singularity (or how it negotiates into a bounce), we need some  basic knowledge of the BKL-M behavior.  The most general  solutions to the Einstein equation near and toward the cosmological singularity  of the Kasner type (not the Friedmann class where matter does matter) is the inhomogeneous Kasner universe (with two directions contracting and one direction expanding), interspersed with the mixmaster behavior, i.e.,  switching of the direction of expansion from one axis to another  in very short time intervals.   This so-called ``inhomogeneous mixmaster solution" represents the generic behavior of the universe toward the cosmological singularity. This result has been reworked by many authors employing different methods, e.g.,  Uggla et al \cite{Uggla}   constructed an invariant set which   forms the local past attractor for the evolution equations, a framework  for proving rigorous theorems concerning the asymptotic behavior of spatially inhomogeneous cosmological models. A more direct  approach relies on methods in dynamical systems \cite{HeiUgg}.  The BKL-M behavior  is also supported by Garfinkle's  numerical simulations  in vacuum spacetimes with  no symmetries \cite{Garfinkle}.

Three main features in the approach to cosmological singularity are, therefore:   1) "Matter doesn't matter" in the BKL-Kasner classes of extrinsic curvature (`velocity')- dominated spacetimes\footnote{An important exception is the case when a massless scalar  field is present, which obeys an equation of state $p = \rho$ like a stiff fluid. 
As is rigorously shown in \cite{AndRen}, such a scalar field will suppress the BKL oscillations  during the evolution towards the singularity.   The relevance of stiff matter in the early universe is noted by Barrow \cite{BarStiff}.} while matter does matter in the FLRW classes. 2) Extrinsic curvature (from expansion or contraction)  dominates over the intrinsic (3-geometry) curvature, except in brief intervals, which are known as the `mixmaster bounces' (the universe point bouncing against the  3-fold symmetric potential walls, receding during contraction) in mini-superspace;  3) Inhomogeneous mixmaster behavior means that  every point on the 3-geometry undergoes an independent mixmaster dynamics. 

We begin with some basics of {\bf Bianchi cosmology}. 
 Bianchi classified all spatially homogeneous spacetimes into nine types \cite{Bianchi} (some types are further distinguished by a sub $0$ or $h$, such as Type $\mathrm{VII}_{h}$, etc.) \cite{RyaShe}. The class of Bianchi Types $\mathrm{I}$--$\mathrm{IX}$ spaces,
especially the Type $\mathrm{I}$ and Type $\mathrm{IX}$ spaces, encompasses many important classical cosmological models for the early universe.  

The line elements of the Bianchi universes are given by
\begin{align}
ds^{2} = -dt^{2} + \sum^{3}_{a,b=1} \gamma_{ab}(t)\sigma^{a}(x)\sigma^{b}(x),
\label{IRST1}
\end{align}
where $\gamma_{ab}(t)$ is the metric tensor and $\sigma^{a}(x)$ are the invariant basis one-forms on the homogeneous hypersurfaces, satisfying the structure condition $d\sigma^{a} = \frac{1}{2} C^{a}_{bc}\sigma^{b} \wedge \sigma^{c}$, where $C^{a}_{bc}$ are the structure constants of the underlying Lie group. For a diagonal type-$\mathrm{IX}$ (mixmaster)  universe, $C^{a}_{bc} = \epsilon_{abc}$ the totally \nobreak antisymmetric tensor, and
$\gamma_{ab}=l_{a}^{2}\delta_{ab}$, where $l_{a}$ are the three principal radii of curvature. The case when two of the three $l_{a}$'s are equal is called the {Taub universe}, the case when all three are equal gives the closed Friedmann--Lema\^{\i}tre--Robertson--Walker (FLRW) universe .

The nonrotating, diagonal {\bf mixmaster universe} with three principal radii of curvature  $\ell_{a (=1, 2, 3)} $ form a  minisuperspace \cite{mix}
which can be  represented pictorially  by the shape $\theta$ and deformation $\b$ parameters,   related to $\ell_{a}$  by
\be
\ell_{1, 2}= \ell_0 \exp[\b \cos (\theta \mp \pi/3)], \;\;\; \ell_3 = \ell_0 \exp (- \b \cos \theta)].
\ee
where $\ell_0^3 \equiv \ell_1 \ell_2 \ell_3$. With threefold symmetry, the origin of this space gives the FLRW universe, while points along the three axes ($\theta=0, 120, 240^o$) constitute the family of the Taub universes. Along any axis, say $\theta=0$,  increasing $\beta$ in the positive sense gives an "oblate" configuration while negative $\b$ gives a "prolate" configuration. The dynamics of the mixmaster universe is depicted by trajectories in this minisuperspace. The velocity of the universe point is determined by the extrinsic curvature, with free motion depicting the { Kasner universe}. The influence of spatial anisotropy (intrinsic curvature) is depicted by a set of moving potential walls. The universe point bouncing off these
walls signifies a shifting in the contracting Kasner axis,
while wandering into one of the three corners signifies an
oscillating and "flattening pancake" (oblate) configuration
There is also one situation when the universe point
bounces off the three walls at a special incidence angle so
that it can continue to hover loosely around the origin.
These characteristic regimes of the mixmaster universe are
called, respectively, (a) the "bounce" (b) the "channel
run" and (c) the "quasi-isotropic" solutions. These solutions
have the approximate geometric configurations as
the three limiting cases of the Taub universe: i.e., (a)
$\b>0$, (b) $\b<0$, and (c) $\b= 0$, respectively.
For any massless field in a spacetime which is a vacuum
solution to Einstein's equations R=0, quantum effects of
curvature are important irrespective of the type of coupling.
 
The {Bianchi Type $\mathrm{I}$ universe}  has zero intrinsic curvature but nonzero extrinsic curvature from its expansion and contraction, whose rates differ in three directions. The line element is given by
\begin{align}
ds^{2} = - dt^{2} + \sum_{i,j=1}^{3} \ell^{2}_{ij}(t) dx^{i}dx^{j} = a^{2}(\eta)\Bigl[-d\eta^{2} + \sum_{i,j=1}^{3} e^{2\beta(\eta)_{ij}}dx^{i}dx^{j}\Bigr]
\label{BRHH1}
\end{align}
where $\eta = \int dt/a$ is the conformal time and $\beta _{ij} $ is a symmetric, traceless $3\times3$ matrix describing the anisotropy of the geometry. For Bianchi Type I one can choose a coordinate such that $\beta _{ij}$ is diagonal\footnote{This diagonal form has full generality because there is no spatial curvature. In a Bianchi Type IX universe where the spatial curvature is present, spacetimes represented by the full matrix $\beta _{ij}$ are more general than that of the diagonal metric which is Misner's mixmaster universe \cite{mix}; the off-diagonal components signify rotation \cite{Ryan72}.} with the three elements $\beta_{i}$. The line element of Bianchi Type I universe is thus
\begin{align}
ds^{2} = -dt^{2} + \sum_{i=1}^{3} \ell^{2}_{i}(t)(dx^{i})^{2} \label{BRHP1}
\end{align}
where $\ell_{i}=a(t)e^{\beta_{i}}$ is the scale factor in the
$x_{i}$ direction. This is a generalization of the spatially flat
isotropic FLRW universe to the case where the universe expands anisotropically. 
A useful quantity for this measure is the anisotropy parameter 
$Q_\beta \equiv - \ha \sum_{i>j}(\dot \ell_i / \ell_i -\dot \ell_j/ \ell_j)^2$, where an overdot denotes $d/dt$.
When there is no matter present a solution to the
vacuum Einstein equation exists, called the {\bf Kasner universe}:
\begin{align}
\ell_{i}(t) = t^{p_{i}}, \quad {\rm where} \quad \sum_{i=1}^{3}
p_{i}^{2} = \sum_{i=1}^{3} p_{i} =1.\label{ch03:eqn3.57}
\end{align}
We see from this relation that the universe expands when two of the $p_{i}$s are positive and one negative, and contracts when two of the $p_{i}$s are negative and one is positive. The Kasner solution is important because it is a generic behavior of the universe at every point in space near the singularity where the most general solution of the Einstein equation 
\cite{BKL} is found to be an inhomogeneous Kasner solution. It is also known as a `velocity-dominated solution'  \cite{ELSachs} reflecting the fact
that near the cosmological singularity the extrinsic curvature
(measuring the time rate of change of the scale factors) dominates over the intrinsic curvature (in the 3- geometry), thus the saying ``matter doesn't matter'' in the BKLM generalized Kasner class of solutions near the cosmological singularity, in contradistinction to the FLRW class, which requires the presence of matter.  Defining the isotropic expansion rate $\alpha = \ln a$ then we see $p_{i}=\alpha + \beta_{i}$. The spatially flat FLRW universe corresponds to the case $\beta = 0$ with scale factor \break $a=e^{\alpha}$.

\subsubsection{Weyl Curvature}

For an analysis of the Weyl curvature tensor in   spatially-homogeneous cosmological models of classical general relativity, see Barrow and Hervik 2002 \cite{BarHer}
For a review of the entropy of universe with gravitational and thermal contributions,  the entropy of black holes and cosmic horizons entropy , see,  Grøn 2012 \cite{Gron12} 
 

\subsubsection{Gravitational Entropy: Weyl and Other measures}  

Assuming the interested reader would have read Penrose's original paper on the WCH, we will not repeat how he defines the gravitationl entropy via the Weyl curvature. Instead, to expand the horizon somewhat we mention a few alternative definitions of gravitational entropy.\\

\noindent {\bf Kinematic vs Weyl singularity}:   

Lim et al 2007 \cite{Lim} and Coley et al 2009 \cite{Coley} studied what they called the `kinematic singularities', namely,    the Hubble expansion scalar, the acceleration vector,  the shear tensor and the vorticity tensor in the decomposition of the covariant derivative of the 4-velocity of a congruence  of worldlines.   \\

\noindent{\bf Bel–Robinson tensor}

Clifton, Ellis and Tavakol  \cite{Clifton} proposed a measure of gravitational 
entropy based on the square-root of the Bel–Robinson tensor of free gravitational fields,  which is the unique totally symmetric traceless tensor constructed from the Weyl tensor.    (For the square-root to exist as a unique factorization of the Bel–Robinson,  the spacetime is required  to contain gravitational fields that are only Coulomb-like or only wavelike.) These authors show that their entropy measure i) is in keeping with Penrose’s Weyl curvature hypothesis,  ii) reduces to the usual Bekenstein–Hawking value measure  for Schwarzschild black hole and  iii) evolves like the Hubble weighted anisotropy of the free gravitational field  for scalar perturbations of a Robertson–Walker geometry thus increases as structure formation occurs, and  iv)  for the  Lema\^{\i}tre Tolman-Bondi inhomogeneous cosmological models,  they found conditions under which the entropy increases.  \\

 \noindent {\bf Kullback-Leibler relative information entropy}

While most works try to place the gravitational entropy defined by the Weyl curvature in relation to thermodynamic entropy,   Hosoya,  Buchert and Morita \cite{Hosoya} try to relate it to information entropy.  They introduced the Kullback-Leibler relative information entropy to express the distinguishability of the local inhomogeneous mass density field from its spatial average on arbitrary compact domains.    Comparing  the Kullback–Leibler entropy to the Weyl   curvature tensor invariants.  Li et al \cite{LiNan} calculated these two measures by perturbing the standard cosmology.  Up to the second order in  the deviations from a homogeneous and isotropic spacetime  they find that they are correlated and can be linked via the kinematical backreaction of a spatially averaged universe model.

\subsection{Semiclassical Gravity: Entropy of Gravitons and WCH}

The Weyl curvature tensor measures the anisotropy  and inhomogeneities of spacetime.  Since it is the radiative sector of the Riemann tensor, it measures  the curvature associated with gravitational waves, which, of course, are the weakly inhomogeneous perturbations off a background spacetime.   Quantizing the linear perturbations from a background gives rise to gravitons. The short wavelength sector under the Brill-Hartle-Isaacson average  behaves like radiation fluid with the same equation of state as  photons, namely,  $p=\rho/3$, with  $p$ the pressure and $\rho$ the energy density.   Gravitons, like photons,  have two polarizations but instead of spin-1 they are spin-2 particles.  For these reasons gravitons are the easiest entry point to see the connection between WCH and  gravitational entropy. 

By semiclassical we mean a quantum field, here, the graviton spin-2 field, propagating in a classical background spacetime. 
(We also distinguish between quantum field theory in curved spacetime,  under the test field situation where the background is prescribed, and  semiclassical gravity where the background spacetime and the quantum field are determined in a self-consistent manner).  By quantum we mean both the matter field and the spacetime are quantized.  We reserve the term `quantum gravity'  to refer to theories about the microscopic structures of spacetime at the Planck energy. We will explain what quantum cosmology means in individual cases later, e.g., quantum cosmology of the mini-superspace as in \cite{mss}  refers to the situation where the 3- geometry is quantized, regardless of whether the spacetime is an emergent entity or a fundamental one.   However,  `loop quantum  cosmology'  or `string cosmology'  is trickier, because `quantum' could mean   quantizing the metric or the connection of GR, and if GR is an effective low energy theory,  the quantized collective excitations are like phonons, a far cry from the fundamental constituents, or  the `atoms',  of spacetime.   It could also mean the cosmology (after the Planck time) derived from a more fundamental theory of spacetime based on strings, loops, sets or simplicies.  Explaining how the spacetime manifold emerges from the interaction of these fundamental constituents is a highly nontrivial task, probably the most severe challenge facing the proponents of these theories of quantum gravity \cite{E/QG}.   
We shall return to this point  in the concluding remarks. 

\subsubsection{Graviton backreaction in FLRW universes}  

We consider this issue at two levels:  1) At the quantum field theory in curved spacetime (QFTCST) \cite{BirDav,FullingBook,WaldBook,ParToms} or test field level, we ask,  under what conditions  would there be particle creation from the vacuum.  It is easy to see that there is no particle production for a massless conformally coupled field in an isotropic and homogeneous universe, because such a spacetime is conformally-flat. But graviton production can happen.  This is because each of the two polarizations of the graviton behaves like a massless minimally coupled scalar field \cite{Gri74,ForPar77}.  Graviton production and its effects have been studied by many authors \cite{Gri75,HuPar77,Hartle77}.  As far as the {\it matter} entropy budget is concerned,  as   mentioned above, with gravitons acting like photons (notwithstanding the big difference in their scattering cross sections), their  production increases the thermal entropy of the matter content. 

 2) To see how this is related to the {\it gravitational} entropy measured by the Weyl curvature tensor,  
one needs to find out the backreaction of  matter fields on  the dynamics of spacetime. This entails finding a  self-consistent solution of  the equations of motion for both the matter fields and the spacetime they live in, namely, the semiclassical Einstein equation.  This backreaction constitutes a channel where the energy and entropy of spacetime can be related to that of the quantum fields, with particle creation acting like a `transducer' between the gravity and matter sectors.  This way of thinking was first proposed by Hu in 1983 \cite{HuPLA83},  having worked out the anisotropy damping scenario due to conformal massless particle creation in Bianchi type I universes earlier with Hartle and Parker.  We shall discuss these results in Sec. IV.    

Here, with the FLRW universe, the problem is simpler, since the background spacetime is not only homogenous but also isotropic.  In the beginning the Weyl curvature tensor is nonzero because of the perturbations, the  gravitational waves or the gravitons,  imparting the universe with some gravitational entropy. At the end, after graviton production ceases,  the spacetime returns to the FLRW universe with zero Weyl curvature tensor.      One can see that the increase of matter entropy from the production process is accompanied by  a decrease of gravitational entropy. 
This is discussed in \cite{NesOtt,Nesteruk}. 
 
Now,  what causes the graviton production to cease? The answer comes from  solving the semiclassical Einstein equations. Doing so,  Hu and Parker 77 \cite{HuPar77} show that the backreaction of created gravitons leads to a change in the power expansion law $a = t^q$ towards $q = 1/2$, i.e., radiation dominated.  And a radiation-dominated RW universe admits no particle creation because the time-dependent frequency of the normal modes of the quantum field contains a term $d^2a/d\eta^2$ where $\eta$ is the conformal time. A radiation-dominated RW universe behaves like $a = \eta$, rendering  $d^2a/d\eta^2=0$.  Thus,  after a certain relaxation time the behavior of the gravitons created corresponds to the evolution of classical radiation in an isotropic universe.   With backreaction we see that as the gravitational entropy density goes to zero, graviton production will cease and no further backreaction will take place. 

The fact that the backrection of graviton production can alter the equation of state of matter was pointed out by Grishchuk in 1975 \cite{Gri75} and shown with field theory calculations by Hu and Parker \cite{HuPar77} and Hartle \cite{Hartle77}.  We will continue this description of the backreaction of particle creation on background spacetime dynamics in  Sec. V and discuss how the damping of anisotropy and inhomogeneities of spacetime bears on gravitational entropy and the WCH. 

\subsubsection{Entropy of Gravitons in the Gowdy Universe}

Husain \cite{Husain} paraphrased  Penrose's ideas of gravitational entropy based on the WCH into three conditions:  ``If the Weyl curvature is to be identified with entropy, it should satisfy the following criteria that are normally associated with entropy.
(l) It should increase monotonically with respect to a
time coordinate from the initial singularity where it is
finite or zero.
(2) Its value should increase with the number of quanta
of the gravitational field. 
(3) Clumped configurations of matter or gravitational
quanta should correspond to higher Weyl curvature
values than unclumped configurations."

Conditions (1) and (2)  are illustrated in the example above, regarding graviton production and its backreaction on the spacetime in a FLRW universe.  For conditions (2) and (3), we can see an example for the Gowdy universe \cite{Gowdy}.  The Gowdy three-torus ($T^3$) universe is an exact solution to the vacuum Einstein equations interpreted to be a single polarization of gravitational waves propagating in an anisotropic, spatially inhomogeneous background.

Appying Hamiltonian quantization via the Arnowitt-Deser-Misner (ADM) method Berger \cite{Berger}  studied  the production of gravitons from empty space. 
At large times, graviton number is well defined since the solution is in WKB form.  The creation process produces the anisotropic collisionless classical radiation.
Near the singularity, the model behaves like an empty Bianchi Type I universe at each point in space (locally Kasner). 

Beginning with the square of the curvature as an operator, Hussain calculated its expectation values in states of clumped and unclumped gravitons.  These results led him to conclude  that the curvature contains information about the entropy of the gravitational field in this class of quantum cosmology models.

Finally,  in the context of quantum cosmology based on the Wheeler-DeWitt (WdW) equation we mention the work of
Grøn and  Hervik \cite{GroHer}  These authors  introduced a measure  of ‘gravitational entropy’ behaving in accordance with the second law of thermodynamics and investigated  its evolution   in Bianchi type I and Lema\^{\i}tre–Tolman universe models.  They work with the quantity $\Pi \equiv P \sqrt h = \sqrt h (\frac{C^2}{R^2})^{\ha}$ which is the ratio  between the Weyl tensor squared $C^2$ and the Ricci tensor squared $R^2$,    integrated over $dV = \sqrt h d^3x$ in a comoving coordinate $x^i$, where $h$ is the 3-volume element. They also considered the expectation value of the Weyl curvature squared and the Ricci curvature squared, taken with respect to the wave functions of the universe satisfying the Wheeler-DeWitt equation.  They then investigated whether a quantum calculation of initial conditions for the universe based upon the Wheeler–DeWitt equation supports Penrose’s Weyl curvature conjecture or not for  the Bianchi type I universe models with dust and a cosmological constant and the Lema\^{\i}tre–Tolman universe models.  They investigated two versions of the hypothesis. The local version based on $P^2$ fails to support the conjecture whereas  a non-local entity based on $\Pi^2$  showed more promise concerning the conjecture. 
Grøn and  Hervik commented that their findings seem to be corroborated by that of  Pelvas and Lake  \cite{PelLak} on the one hand, who showed that  local entities like $P^2$ cannot be a measure of gravitational entropy, and on the other hand,  by that of Rothman and Anninos \cite{Rothman},  who showed that a nonlocal quantity they constructed using the volume of the phase space could.   The investigations of these authors seem to suggest that while one  cannot say that the Weyl tensor is directly linked to the gravitational entropy, a certain non-local entity which is constructed from it has an entropic behavior, and reflects the tendency of the gravitational field to produce inhomogeneities. 

We shall say some more about ADM and WdW, and about loop quantum cosmology  in Sec. III.  But 
we will not discuss quantum geometry and gravitational entropy (see, e.g., \cite{Balasub})  applied to black holes or in the  much studied holographic entanglement entropy vein \cite{RyuTaka}  for spacetimes with boundaries. 

\section{Cyclic Cosmology and Trace-Anomaly Induced Bounces}

\subsection{Penrose's conformal cyclic cosmology (CCC)}

This theory is Penrose's recent favorite in terms of cosmological models -- favorite in that after his 2005 proposal
he has gathered his thoughts on  entropy, black holes, time and cosmology in a 300 page 2010 book \cite{CoT} which dwells on these foundational issues of Nature in a lucid and inspiring way.    In his more recent 2017 book \cite{FFF} of over 500 pages,  ``Fashion, Faith, and Fantasy in the New Physics of the Universe"  Penrose refers  (Sec. 4.3) to his  "conformal cyclic cosmology" as an idea so fantastic that it could be called "conformal crazy cosmology"\footnote{An element of craziness needs to creep in before an idea ascends to the order of the three Fs:  Fashion, Faith, and Fantasy.  I read this  as a warning of an enlightened guru speaking to his present adherents and future believers: Safeguard your independent thinking before you become completely converted. }.

To see what CCC is, the reader may benefit from reading just 4  pages of exposition in \cite{Penrose2006}.  For the  background  it is useful to know something about singularities in GR, e.g., in \cite{Geroch},  and in particular, the class of isotropic cosmological singularities, where the Weyl tensor is finite or preferably zero in the WCH. At the isotropic singularity even though the Ricci tensor related to matter may approach infinity, the Weyl curvature remains finite.   Goode and Wainwright \cite{GooWai,Goode} proposed a definition of isotropic cosmological singularities which is adopted by Tod \cite{Tod} whose work Penrose refers to.  The essential idea  is that the physical metric which is singular at cosmic time $T=0$,  is conformally related to a metric which is regular at $T = 0$, i.e. the singularity arises solely due to the vanishing of the conformal factor.  The conformal transformation which is used to define  an isotropic singularity is an essential ingredient in Penrose's conformal cyclic cosmology.  Namely, the Big Bang can be conformally represented as a smooth spacelike 3-surface, across which the space-time is, in principle, extendable into the past in a conformally smooth way.


According to Penrose \cite{Penrose2018}, the Big Bang (BB)  of our `aeon' is the {\it conformal continuation} of the remote exponential expansion of a previous aeon.  There is a pre-BB aeon, and after our future Big Crunch (BC) -- actually Penrose prefers to see a long exponential expansion (EE) to our future -- there is a post-BC or post-EE aeon.   In this sense  a cyclic cosmology scenario naturally emerges,  where  the aeons are linked by conformal transformations.    

\subsection{Cyclic cosmology from particle/ string models. Big crunch}

 No classical matter can exist with a pressure $p$ to energy density $\rho$ ratio $w \equiv p/\rho >1$.  The stiffest Zel'dovich equation of state with $w=1$ is the upper limit of classical matter because that is when the speed of sound equals the speed of light.    
Quantum matter can break this limit, e.g., a scalar field which drives inflation has an effective equation of state with $w=-1$.    
In the recent two decades  cosmological models with a scalar field providing an effective equation of state $w>1$  have drawn some  interest,  motivated by the possibility of cyclic cosmologies based on string theory- related models, notably the  Pre-Big Bang model of Gasperini and Veneziano  \cite{GasVen} and the ekpyrotic cosmologies of Steinhardt and Turok \cite{SteTur}. The designers of these theories want to see the universe  contract towards a big crunch singularity in a smooth  homogeneous and isotropic (FLRW) manner, so it can be reborn in an equally smooth manner to be able to sustain a large number of cycles, ideally ad infinitum.  

\subsubsection{Cyclic and Ekpyrotic Cosmologies} 

Ekpyrotic cosmologies solve  the flatness problem in the contraction phase in a way similar to how inflation solves it in the expansion phase.  Both are looking for the conditions where the time rate of change of the scale factor (Hubble) becomes close to a constant for a period of time (around 65 e-folding time to be compatible with the entropy contents of the observed universe).  It is easy to see this from two simple equations, the Einstein $G_{00}$ equation for the dynamics of spacetime and the continuity equation (divergence-free condition) for matter. 
In a FLRW universe with scale factor $a$ and Hubble expansion rate $H = \dot a/a$, where an overdot denotes taking the derivative with respect to cosmic time,  the Einstein $G_{00} = 8 \pi G \rho$ equation reads:  $ H^2 = - k /a^2 + \frac{8 \pi G}{3}\rho$, where $k =  0, 1, -1$ correspond to the flat, closed and open universes and $\rho$, the total energy density, contains different kinds of classical matter and quantum fields, represented by equations of state with different values for $w \equiv p/\rho$. The familiar cases of $w = 0, 1/3$ in classical cosmology are for dust and radiation, respectively.   We shall first show i) why $w= -1$ for inflaton field, and then explain ii) how a special kind of matter with an equation of state $w >1$ affects the Hubble rate, and what it does for the ekpyrotic cosmologies.  
 
It is well-known how inflation solves the flatness problem, namely,  by assuming that the potential $V(\phi)$ of the inflaton field  $\phi$ is sufficiently flat so that it  rolls down the slope very slowly.  That imparts a constant energy density in the Einstein equation which dominates over the other sources, which all drop as power laws of the scale factor during an expansion phase.  The pressure-energy density ratio for a quantum field is $w\equiv p/\rho =  \frac{ \ha  {\dot\phi}^2-V}{\ha  {\dot\phi}^2 + V}$.  When $\ha  {\dot\phi}^2 \ll V$ as in slow-roll inflation, $w \rightarrow -1$.   

From the continuity equation we  get $\rho \propto a^{-3(1+w)}$,  namely, for non-relativistic matter $ \rho_{nrel} \propto a^{-3}$, for radiation $\rho_{rad} \propto a^{-4}$, for curvature anisotropy $\rho_{anis} \propto a^{-6}$, and for some  `ultra-stiff' nonclassical matter with $w>1$,  $\rho_{ust} \propto a^{-r}$ where $r > 6$.   Thus,  during contraction, as $a$ gets smaller, contributions from this ultra-stiff matter with $w>1$ will dominate over other forms of energy density sources in governing the expansion rate.   

This is why  the  cyclic and ekpyrotic cosmological models focus on the Big Crunch singularity and why scalar fields with equations of state $w>1$ are of special interest.  This `ultra-stiff' matter in a homogeneous and isotropic universe can be realized by a scalar field  $\phi$ with a steep negative potential energy  $V(\phi)< 0$.   The wits and skills in designing the  potential of a quantum field to fit one's special needs --  these activities which dominate the theoretical stage in inflationary cosmology -- will also be active for this newer vein of cosmology.  We refer interested readers to the nice review of Lehners \cite{Lehners} from which they can trace back to the original papers.


Of special interest to our concern  is whether the mixmaster behavior might be altered by the effect of a scalar field with equation of state $w>1$.  These authors of Ref.  \cite{Erickson} show that if $w> 1$, the chaotic mixmaster oscillations due to anisotropy (extrinsic curvature) and (intrinsic) curvature are suppressed, and the contraction is described by a homogeneous and isotropic Friedmann equation.  
Referring to string theory related models,  they show that chaotic oscillations are suppressed in  $Z_2$ orbifold compactification and contracting heterotic M-theory models if $w>1$ at the crunch.

However, as pointed out by Barrow and Yamamoto \cite{BarYam} this result stands only when the ultrastiff pressures are  isotropic. 
  The inclusion of simple anisotropic pressures  when pressures can exceed the energy density stops the isotropic Friedmann universe from being a stable attractor as an initial or final singularity is approached.  Thus the situation with isotropic pressures, studied earlier in the context of cyclic and ekpyrotic cosmologies, is not generic, and Kasner-like behavior occurs when simple pressure anisotropies are present.

Including the consideration of stochastic effects, as invoked in Linde's `chaotic' inflation or Starobinsky's stochastic inflation scenarios \cite{EtInf,StoInf},  the effective scalar field can climb up  the potential in some regions of space, which leads to an increase in the energy density.  However, as Brandenberger et al \cite{BranBkrn} show,  the backreaction of fluctuations that have already exited the Hubble radius will lead to a decrease in the effective energy density,  which is strong enough to prevent eternal expansion in the late-time dark-energy phase of ekpyrotic cosmology.  

\subsubsection{Big Crunch with Dark Energy in Semiclassical Cosmology}

As was mentioned above, how the universe evolves in the future-- whether it approaches a  singularity (Big Crunch),  or it can avoid it (a Bounce), and how it approaches the singularity or the minimal radius,  are important factors to consider in any cyclic cosmology proposal.  Without assuming any exotic matter,  Kennedy et al \cite{Kennedy} studied the influence of the vacuum expectation value of the energy-momentum tensor of a conformally-invariant free quantized field without particle creation on the spherically symmetric gravitational collapse of a dust cloud. Using the same method as used by  Fischetti et al \cite{FHH79} and Anderson \cite{And83}  for the early universe,  they found qualitatively similar results, namely,  for certain values of an arbitrary parameter the collapse proceeds to singularity, although in a manner which differs from the classical (Oppenheimer-Snyder) behavior; for other values of the parameter the collapse terminates before singularity and thereafter the dust cloud expands. 

Ever since the discovery of a sizable fraction (0.7) of dark energy in the cosmological matter-energy contents, theories about the future of our universe need to take into account the bounds provided by stringent cosmological data.  The combined 2018 Planck data together with other major observations give the dark-energy equation of state parameter  $w_{de} = -1.03 \pm 0.03$, consistent with a cosmological constant ($w=-1$).  One can regard this as weakly favoring `phantom' dark energy ($w <-1$), which can lead to a future singularity called `Big Rip' \cite{Caldwell}.  (A rather complex classification of different possible future singularities is given in, e. g., Ref. \cite{Carlson} Sec. IA. )  Assuming that the dark energy which causes these singularities to behave like a perfect fluid,  Carlson et al \cite{Carlson} studied the effects which quantum fields and an $\alpha_0R^2$ term in the gravitational Lagrangian have on future singularities. Special emphasis was placed on those values of $\alpha_0$ which are compatible with the universe having undergone Starobinsky inflation \cite{Star80} in the past,  the same values which allow for stable solutions to the semiclassical backreaction equations in the present universe.

\subsection{Semiclassical Gravity: Trace anomaly induced bounces}

For conformally- invariant fields in conformally- flat spacetimes in the conformal vacuum state, the regularized stress energy tensor is given by    \cite{And83} 
\bea \label{Tab}
<0|\hat T_{ab} |0> &=& \frac{\a}{3} (g_{ab}{R^{;c}}_{;c} - R_{;ab} + R R_{ab} - \frac{1}{4} g_{ab}R^2) \\ \nn
&+& \b(\frac{2}{3}R R_{ab} - {R_a}^c R_{bc}+\ha g_{ab}R_{cd}R^{cd}-\frac{1}{4}g_{ab}R^2),
\eea
where $R_{ab}$ is the Ricci tensor, $R$ is the scalar
curvature, and $\a$ and $\b$ are constants which depend on the number and types of fields present. Dimensional regularization gives 
\be
\a = \frac{1}{2880\pi^2}(N_S +6N_\nu+ 12 N_V); \;\; \b= \frac{1}{2880\pi^2}(N_S+11N_\nu+62N_V)
\ee 
where $N_S, N_\nu, N_V$ are respectively  the number of scalar fields,  four-component neutrino fields, and  Maxwell fields included in one's theory.

Taking the trace of  \eqref{Tab} we see that it is non-vanishing even though for conformal fields the classical stress energy tensor is traceless.  This is why it is called the trace anomaly.  The semiclassical Einstein equations have been solved for the effects of backreaction in the 70s in various cases by the following authors: Ruzmaikina \& Ruzmaikin \cite{Ruz2} were the first to work on this problem, they,  and later,  Gurovich  \&  Starobinsky \cite{GurSta}  studied the case $\b=0$, Wald \cite{Wald}  investigated the case $\a =0$, and Fischetti, Hartle, and Hu \cite{FHH79}  the cases $\a \le 0$, all values of $\b$ and $\a >0, \b <3\a$.  Each of these was done for
spatially-flat FLRW spacetimes containing classical radiation
and having no cosmological constant. Starobinsky  \cite{Sta80}
investigated the case $\a<0, \b>0$ for FLRW
spacetimes with no classical radiation or matter and
no cosmological constant. 

The most complete investigation of  the back-reaction problem for conformally-invariant free quantum fields in spatially-flat homogeneous and isotropic spacetimes containing classical radiation was carried out by Anderson \cite{And83}.  He considered only solutions which at late times approach the appropriate solution to the field equations of general relativity, the so-called asymptotically classical solutions,  with the following results:  For all  $\a, \b$ with  $\a> 0$  there are many such solutions, while for all  $\a,\b$  with  $\a <0$  there is only one such solution. For  $\b > 3 \a > 0$, there is always one solution which undergoes a "time-symmetric bounce" and which contains no singularities or particle horizons. For $\a>0$  there is always at least one solution which begins with an initial singularity and which has no particle horizons. For all  $\a, \b$  there is always at least one solution which begins with an initial singularity.  

In a sequel paper \cite{And84}, Anderson added the consideration of  nonzero spatial curvatures and/or a nonzero cosmological constant. The qualitative behaviors are the same as for spatially-flat spacetimes with zero cosmological constant.  Instead, for  $\b > 3 \a > 0$, there is always one solution which undergoes a "time-symmetric bounce" and which contains no singularities or particle horizons. The differences caused by the spatial curvature and cosmological constant include the initial behavior of the time-symmetric bounce solution and, if the spatial curvature is nonzero, the initial behavior of many solutions for the cases $\b \approx 3\a>0$   and $\b \approx 3\a <0$.  For further work on semiclassical backreaction of conformal quantum fields in a  universe with a cosmological constant, see, e.g., Ref.  \cite{AzuWad}.

\section{Singularity Avoidance in  Quantum Cosmology}

Readers interested in having a broader perspective of this issue   might want to first learn of the major milestones in the developments of quantum gravity.  A summary is given by C. Rovelli \cite{RovMilesQG}.  

The first stage of quantum gravity began with a series of papers by Bergmann pursuing  a program concerned with the quantization of field theories which are covariant with respect to general coordinate transformations, like the general theory of relativity. All these theories share the property that the existence and form of the equations of motion is a direct consequence of the covariant character of the equations.  Dirac's work  \cite{DiracQG} on  the quantization of constrained systems is a cornerstone in this pursuit.
A more sophisticated and systematic treatment can be found in the book by Henneaux and Teitelboim \cite{HenTei}.\\

\noindent {\bf ADM Quantization}.  As most researchers would agree, the canon of canonical quantization of general relativity is the ADM formalism for quantizing 3-geometries in a  3+1 (space-time) decomposition, developed by Arnowitt, Deser and Misner in a series of papers 1959-61. See their summary in \cite{ADM62}.  Read also Kucha\v{r} \cite{Kuchar},  Ashtekar \cite{AshQG},  the monographs of Rovelli \cite{RovQG} and Thiemann \cite{ThiQGR}. \\

\noindent {\bf Wheeler-DeWitt equation}.

The Wheeler-DeWitt equation \cite{WheDeW} is a Hamilton-Jacobi equation formulation of  the ADM quantization on the superspace, the space of all 3-geometries.  (For a short introduction with some historical tidbits, read \cite{RovStrange}.)   Minisuperspace refers to the truncation of the infinite dimensional superspace to a few dimensions, such as the scale factors of the Bianchi universes.  Quantum cosmology was studied by Misner \cite{mix,mss} by applying the ADM quantization to the mini-superspaces of Bianchi Type I and IX universes.    Read also Ryan \cite{Ryan72}. 

The second wave of quantum cosmology arrived in the mid-80s with the considerations of the initial conditions for the wave functions of the universe, notably,  Hartle and Hawking's no boundary condition \cite{HarHawNB} and  Vilenkin's tunneling wave function \cite{Vilenkin}.  That was also the time when the two major programs of quantum gravity took off, namely,  superstring theory \cite{stringBooks}  and loop QG \cite{LQG} (and its modern version of spin foam network \cite{SpinFoam} which has a more obvious microscopic appeal). Supersymmetric quantum cosmology is well represented by the monographs of D'Eath \cite{DEathBook} and of Moniz \cite{MonizBook}.  Other major approaches to quantum gravity are nicely represented in Oriti's edited book \cite{Oriti}.  A  state-of-the-art overview of multiple approaches from the words of 37 prominent practitioners is recorded in this most recent book \cite{Armas}.  Please refer to the original papers, reviews and monographs on this subject. 
 
\subsection{Quantum Cosmology: Bounce in Gowdy $T^3$ universe}

Here we summarize Berger's work \cite{Berger74,Berger84}.  The Gowdy $T^3$ universe \cite{Gowdy} can be  interpreted as  consisting of a single polarization of gravitational waves whose amplitudes satisfy a linear wave equation propagating in a non-linear background spacetime.   One can then view these wave amplitudes acting quadratically as a source  on the  first-order equations for  the non-linear  background part of the gravitational field  as a backreaction problem.   Classically the model has an initial “big bang” singularity near which it is velocity-dominated  behaving as a different Kasner solution at each value of the  spatial coordinate orthogonal to the symmetry plane. Far from the singularity, the method of Isaacson \cite{Isaacson} may be used to identify the gravitational waves. Their effective stress-energy tensor is that for a $p = \rho$ fluid in two dimensions propagating in a spatially-homogeneous background spacetime.  At the quantum level,  Berger performed a  canonical quantization of the dynamical degrees of freedom.  An adiabatic vacuum state was introduced and adiabatic regularization used to obtain non-divergent stress-energy tensor vacuum expectation values.  
The regularized expectation value was used as a source for the classical background spacetime in the spirit of semi-classical gravity.  Berger found that  ``The effect of the regularization of the stress-energy tensor expectation value  is to replace the classical singularity with a symmetric bounce. The semi-classical spatially-averaged Gowdy $T^3$ cosmological model thus collapses from a state of large volume  to a minimum volume near $r = 0$  and then re-expands.  The classical singularity is replaced by a symmetric bounce." 

\subsection{Quantum BKL-mixmaster scenarios}

Earlier we mentioned  applications of dynamical system methods and chaos theories to cosmology,  in particular, to the analysis of the BKL-mixmaster behavior \cite{Bogoy,WainEll,Col}.
 For quantum chaos applied to the BKL-M models, we mention the numerical solutions of the Wheeler-DeWitt equation by  Furusawa   \cite{Furu}  and   Monte Carlo simulations of the vacuum Bianchi IX universe by Berger \cite{BergerMC}. For more recent  work we mention two groups of authors on this subject.  The first group's authors are  Bergeron, Czuchry, Gazeau, Małkiewicz and Piechocki. Of particular interest to our theme here is their results on singularity avoidance \cite{Berg15}.
How to get the universe to enter an inflationary stage without soliciting the aid of an inflaton field (cf. Starobinsky inflation \cite{Star80} invokes a scalar field, at least the trace anomaly of it, as was also studied in \cite{FHH79}) is certainly of  interest \cite{Berg16}. (See \cite{HuOCMixInf} for an earlier work on mixmaster inflation via an interacting quantum field.) These authors constructed the quantum mixmaster solutions using  affine and Weyl–Heisenberg covariant integral quantizations.
According to these authors these quantization methods  can  regularize classical singularities which, they claim, the commonly implemented canonical quantization cannot.  From this they show some promising physical features such as the singularity resolution, smooth bouncing, the excitation of anisotropic oscillations and a substantial amount of post-bounce inflation as the backreaction to the latter.  For a recent summary of these authors' work on this subject, see \cite{Berg20}.  

The other group's authors are  Gó\'zd\'z, Kiefer, Kwidzinski, Piechocki, Piontek and G. Plewa.  Kiefer et al \cite{QBKL0}  formulated the criteria of singularity avoidance for general Bianchi class A models in the framework of quantum geometrodynamics based on the Wheeler–DeWitt
equation and give explicit and detailed results for Bianchi I models with and without matter. The singularities in these cases are big bang and big rip. They find that the classical singularities can generally be avoided in these models.  

Gó\'zd\'z et al \cite{QBKL1} use affine coherent states quantization  in the physical phase space and show that  during quantum evolution the expectation values of all considered observables are finite. The classical singularity of the BKL scenario is replaced by the quantum bounce that presents a unitary evolution of the considered gravitational system.    In a later paper \cite{QBKL2} they asked whether their claims may change because  the affine coherent states quantization method  depends  on the choice of the group parametrization.  Using the two simplest parameterizations of the affine group, they  show that qualitative features of their quantum system do not depend on the choice. They maintain that the quantum bounce replacing a singular classical scenario is expected to be a generic feature of the considered system. 

\subsection{Loop Quantum Cosmology (LQC): `Big Bounce'?}

We devote this space to LQC because singularity avoidance is
supposed to be a major accomplishment of LQG \footnote{We could include group field theory (GFT) \cite{GFT} in the present considerations (only) to the extent that  (partial)  equivalence of models employed or results reported can be shown to exist between these two theories \cite{BojoGFT}.  Of course we need to hear  the views of the GFT community on this. }. This is referred to as the `Big Bounce',  in capital letters. 
There are many reviews of  LQC, e.g.,  Ashtekar et al \cite{LQCAsh,LQCAshSin}, Bojowald \cite{LQCBojo}, and a recent book  \cite{LQCBojo}. (Note the loop QG community does not sing in unison  in regard to many major claims in LQC.  Read Ashtekhar and Singh's  2011 status report \cite{LQCAshSin} and the critiques of Bojowald \cite{BojoCritic}.)

Without getting into the detailed claims and critiques we ask two key physical questions: 1) What role would genuine quantum degrees of freedom play in the cosmological singularity proven to exist in classical general relativity?  2) Does quantization of gravity alter the Big Bang?  If it does, will there be a minimum finite scale in the quantum geometry -- the quantum bounce?  (Here,   the canonical formulation of cosmic `time' is often provided by a relational variable such as the scale factor in the `3-volume' or a scalar field.)

There seems to be some consensus  in the LQC community on the   existence of a quantum bounce. 
On the first issue Bojowald asked the question \cite{BojoQBB}: How Quantum is the Big Bang?  Studying  isotropic models with an interacting scalar field he concluded that quantum fluctuations    do not affect the bounce much.  Quantum correlations, however, do play an important role and could even eliminate the bounce. 

 In \cite{BojoDCS}  a new model is studied which describes the quantum behavior of transitions through an isotropic quantum cosmological bounce in loop quantum cosmology sourced by a free and massless scalar field. Using dynamical coherent states Bojowald provided a demonstration that in general quantum
fluctuations before and after the bounce are unrelated. ``Thus, even within this solvable model the condition
of classicality at late times does not imply classicality at early times before the bounce without further
assumptions. Nevertheless, the quantum state does evolve deterministically through the bounce.  These analytical    bouncing solutions corroborate  results from numerical simulations attesting to robustly smooth bounces under the assumption of semiclassicality. " 

There seems to be no controversy in this last statement within  the LQC community.  However, how much of the state of the universe before the bounce is retained, is a subject of debates. \\

\noindent {\bf  Cosmic memory over bounces}

For theorists favoring the bounce scenario, especially the cyclic cosmologies,  an important question to ask is, does the universe retain, after the bounce, its memory about the previous phase.  

Bojowald is of the opinion of `cosmic forgetfulness'  \cite{BojoForget}, 
meaning,  not all the fluctuations (and higher moments) of a state before the bounce can be recovered after the bounce, and values depend very sensitively on the late time state. In \cite{BojoBefore} he showed that quantum fluctuations before the big bang are generically unrelated to those after the big bang.  A reliable determination of pre-big bang quantum fluctuations of geometry would thus require exceedingly precise observations.

Countering this claim,  Corichi and  Singh \cite{CorSin},  in one exactly solvable model,   found that a semiclassical state at late times on one side of the bounce, peaked on a pair of canonically conjugate variables, strongly bounds the fluctuations on the other side, implying semiclassicality. From this result these authors assert that cosmic recall is almost perfect.   See Bojowald's Comments \cite{BojoCom} and Corichi \& Singh's Reply \cite{CorSinRep}.\\

\noindent {\bf  Bianchi I model as a prototype for a cyclical Universe}

For more recent work on this topic, we mention two.    
Montani,  Marchi and Moriconi \cite{MMM} investigated the classical and quantum behavior of a Bianchi I model with stiff matter,  an ultra-relativistic component, and a small negative cosmological constant (at the turnover point).  They apply the Vilenkin wave function, using the volume of the universe, a quasi-classical variable, as the dynamical clock for the pure quantum degrees of freedom.   They found that in a model where the isotropic variable is considered to be on a lattice,  the big-bang singularity is removed at a semiclassical level in favor of a big bounce. The mean value of the Universe anisotropy variables remains finite during the whole evolution across the big bounce,  assuming a value that depends on the initial conditions fixed far from the turning point.   

An even simpler picture is obtained by Wilson-Ewing \cite{Wil-Ewi}: ``In the loop quantum cosmology effective dynamics for the vacuum Bianchi type I and type IX space-times, a non-singular bounce replaces the classical singularity. The bounce can be approximated as an instantaneous transition between two classical vacuum Bianchi I solutions, with simple transition rules relating the solutions before and after the bounce. These transition rules are especially simple when expressed in terms of the Misner variables: the evolution of the mean logarithmic scale factor $\Omega$ is reversed, while the shape parameters $\beta_\pm$ are unaffected. As a result, the loop quantum cosmology
effective dynamics for the vacuum Bianchi IX space-time can be approximated by a sequence of classical vacuum Bianchi I solutions, following the usual Mixmaster transition maps in the classical regime, and undergoing a bounce with this new transition rule in the Planck regime."
At the level of effective dynamics in loop quantum cosmology, for this type of  bounce solutions, Gupt and Singh \cite{GupSin12} 
found the selection rules and underlying conditions for
all allowed and forbidden transitions.\\

\noindent {\bf  Open Issues of LQC}

Bojowald, who first reported a bounce solution in LQC, raised a few cautionary points \cite{BojoNoB}:  a) A conceptual gap exists in the form of  several  unquestioned links  between bounded densities, bouncing volume expectation values, and singularity avoidance, made commonly in arguments in favor of a generic bounce in LQC. b) While bouncing solutions exist and may even be generic within a given quantum representation, they are not generic if quantization ambiguities such as choices of representations are taken into account.  Small-volume BKL behavior in a collapsing universe is shown to be crucially different from the large-volume  behavior exclusively studied so far in LQC. 


\section{Quantum Phase Transition and Quantum Backreaction}

In addition to the popularly known string and loop theories, there is a  handful of  proposals of quantum gravity theories describing sub-Planckian (length scale) physics \cite{Oriti}.  Our present aim is to see  how the WCH fares in these theories, i.e., whether some of these theories predict that the Weyl curvature diminishes by comparison with other components of the Riemann tensor  near the singularity or bounce.  Note that this question may not be answerable by all of them because many of these theories have not yet presented a full account of how  their basic constituents interact,  and at the Planck length and above, give rise to the familiar manifold structure of spacetime described by GR.  What makes this non-trivial and non-straightforward is the possibility of the emergence of   mesoscopic structures between the micro (QG) and the macro (GR), or the occurrence of  phase transitions which, in our opinion, is very likely.  Quantum gravity not being discussed here since it is a topic covered by many books and reviews, we will limit our attention to identifying those micro-theories which predict a robust and smooth (near-isotropic and homogeneous) geometry at the Planck scale rising from sub-Planckian scale by phase transitions or other processes or mechanisms (e.g., polymerization, crystalization).  

\subsection{Quantum Phase Transition: Continuum to Discrete}

\noindent {\bf Asymptotic Safety}

First proposed by Weinberg  \cite{AymSafe}, the asymptotic safety program in quantum gravity \cite{AS}  asserts that  the short-distance behavior of gravity is governed by a nontrivial renormalization group fixed point.  As describd by Bonano and Saueressig \cite{BonSau} 
It provides an elegant mechanism for completing the gravitational force at sub-Planckian scales. At high energies, the fixed point controls the scaling of couplings such that unphysical divergences are absent, while the emergence of classical low-energy physics is linked to a crossover between two renormalization group fixed points. These features make asymptotic safety an attractive framework for the building of a cosmological model.

D’Odorico  and Saueressig \cite{DOdSau}  studied   quantum corrections to the classical Bianchi I and Bianchi IX universes.     The correction terms induce a phase transition in the dynamics of the model, changing the classical, chaotic Kasner oscillations into a uniform approach to a point singularity.   This seems to be consistent with the results obtained from the backreaction of particle creation from quantum fields, namely, the  isotropization of anisotropy, which we mentioned in Sec. 2.3 and which will be described  in Sec. 5.1.\\

\noindent {\bf Causal Dynamical Triangulation CDT}

A systematic and vigorous   treatment of discrete quantum geometry is the Causal Dynamical Triangulations (CDT) program pursued by Ambjorn, Jurkiewicz, Loll et al extensively for decades. CDT  is a lattice model of gravity that has been used to study non-perturbative aspects of quantum gravity.   Read the review of  Loll \cite{Loll},  and  for the most recent development,  Ambjorn et al \cite{AmbJur} :  ``CDT  has a built-in time foliation but is coordinate-independent in the spatial directions. The {\it higher-order phase transitions} observed in the model may be used to define a continuum limit of the lattice theory. Some aspects of the transitions are better studied when the topology of space is toroidal rather than spherical. In addition, a toroidal spatial topology allows us to understand more easily the nature of typical quantum fluctuations of the geometry.  In particular, this topology makes it possible to use massless scalar fields that are solutions to Laplace’s equation with special boundary conditions as coordinates that capture the {\it fractal structure of the quantum geometry}. When such scalar fields are included as dynamical fields in the path integral, they can have a dramatic effect on the geometry." One can view this as a discrete modeling of the Gowdy $T^3$ universe we described in Sec. 3.1. The quantum phase transition is of course the interestingly new and exciting development.  \\

\noindent{\bf Causal Set: Continuum to crystalline phase transition} 

Causal set theory, first  proposed by Sorkin \cite{SorkinCS} and developed extensively by Dowker \cite{DowkerCS} and their associates, has seen significant  developments both in the number of adherents and in scope of its research topics.  How the classical continuum spacetime  as we know it evolved from discrete causal sets is one of the key issues.   Brightwell et al \cite{Bright}  noted that non-perturbative theories of quantum gravity inevitably include configurations that fail to resemble physically reasonable spacetimes at large scales. Yet these configurations are entropically dominant and pose an obstacle to obtaining the desired classical limit. These authors examine this `entropy problem' in a model of causal set quantum gravity corresponding to a discretization of 2D spacetimes. Using results from the theory of partial orders they show that, in the large volume or continuum limit, its partition function is dominated by causal sets which approximate  a region of 2D Minkowski space.  

Surya \cite{Surya12} continued this investigationand presented evidence for a continuum phase in a theory of 2D causal set quantum gravity, which contains a dimensionless non-locality parameter $\epsilon \in (0, 1]$. She also found a phase transition between this continuum phase and a new crystalline phase which is characterized by a set of covariant observables.
For more recent developments, see her review \cite{SuryaLRR} \\

\noindent{\bf Berezenskii-Kosterlitz-Thouless (BKT) transition}

 Antoniadis, Mazur, and Mottola \cite{AnMM} presented a simple argument which determines the critical value of the anomaly coefficient in four dimensional conformal factor quantum gravity, at which a phase transition between a smooth and elongated phase should occur. The argument is based on the contribution of singular configurations (“spikes”) which dominate the partition function in the infrared. The critical value is the analog of c = 1 in the theory of random surfaces, and the phase transition is similar to the Berezenskii-Kosterlitz-Thouless transition. The critical value they obtain is in agreement with the previous canonical analysis of physical states of the conformal factor and may explain why a smooth phase of quantum gravity has not yet been observed in simplicial simulations. \\

\noindent{\bf 2D Quantum Gravity and Random Geometry}

This was an active topic of research in the second half of the 80s. As a representative work,  Gross and Migdal \cite{GroMig}   proposed a nonperturbative definition of two-dimensional quantum gravity, based on a double scaling limit of the random matrix model which they solve  to all orders in the genus expansion. They derived an exact differential equation for the partition function of two-dimensional gravity as a function of the string coupling constant that governs the genus expansion of two-dimensional surfaces. The works  of the principal contributors to this subject  are collected in a book \cite{2DQG}.\\

\subsection{Quantum Backreaction of  Inhomogeneous Superspace Modes}

We have mentioned the backreaction problem earlier.  It may refer to several different effects. We use different symbols to denote them:  (C) the backreaction of inhomogeneous modes of classical gravitational perturbations on the homogeneous modes in classical general relativity, as in the works of Buchert,  Clifton, Ellis et al versus Green \& Wald.  (Q) The  backreaction of the inhomogeneous modes of quantum geometry (in the full superspace of 3-geometries) on the mini-superspace  quantum cosmology as studied in \cite{BruThi,SchThi}. In between the classical and quantum levels lie  the semiclassical and stochastic.  (S) The backreaction in semiclassical and stochastic gravity refers to the effects of  quantum matter field processes, such as the trace anomaly or particle creation, back-reacting on a classical  background geometry:  (SC) Semiclassical gravity when the mean value of the stress energy tensor of matter fields act as the source in the semiclassical Einstein equation; (ST) stochastic gravity when the fluctuations of the quantum fields are  included as sources of the Einstein-Langevin equation. The `quantum backreaction' in the title of this paper refers  to  the middle two levels between the classical and the quantum.     
 
These 4 levels (C-SC-ST-Q) of backreaction are summarized in a recent review with an extensive bibliography \cite{SchThi}. 
These authors start by assessing the question of backreaction, i.e., whether cosmological inhomogeneities have an effect on the large scale evolution of the Universe, focusing on the purely quantum mechanical backreaction. They present one recent approach based on mathematical tools inspired by the Born–Oppenheimer approximation to include backreaction in quantum cosmology.  

We mention three groups of representative work on this issue in the 80s-90s,  then  leave this topic for the readers to explore with the help of this excellent review.   

1) Using path-integral quantization,  Hawking and Halliwell \cite{HalHaw}   assume  that the Universe is in the quantum state defined by a path integral over compact four-metrics.  This can be regarded as a boundary condition for the wave function of the Universe on superspace, the space of all three-metrics and matter field configurations on a three-surface, the same as was proposed by Hartle and Hawking earlier. They
 treated the homogeneous and isotropic degrees of freedom of the Friedmann Universe exactly and the inhomogeneous and anisotropic degrees of freedom in superspace to second order in the perturbations.  For the same model Kiefer \cite{Kie87} calculated explicitly the wave functions for all multipoles of the matter field and for the tensor modes of the metric. D'Eath and Halliwell \cite{DEaHal87} have considered the backreaction of the fermionic perturbations on the homogeneous modes.  Origin of structure in supersymmetric quantum cosmology was studied by Moniz \cite{MonizPRD}. By employing the supersymmetry and Lorentz constraint equations  a set of quantum states was   obtained and  a particular quantum
state which has properties typical of the conventional no-boundary  Hartle-Hawking  solution was identified.

2) In canonical quantum gravity  Kucha\v{r} and Ryan \cite{KuchRyan}  questioned the  validity of physical predictions based on minisuperspace quantization of Einstein's theory of gravitation. They proposed to investigate a hierarchy of models with higher symmetry embedded in models of lesser symmetry,  to spell out the criteria under which minisuperspace quantum results can be expected to make meaningful predictions about full quantum gravity.  As a concrete example they studied a homogeneous, anisotropic cosmological model of higher symmetry (the Taub model) embedded in one of lesser symmetry (the mixmaster model) and  showed that the respective behavior is widely different.  

3) With quantum field theory placed in an open systems setting,  Sinha and Hu \cite{SinHu} used the example of an interacting  $\lambda \Phi^4$ scalar quantum field in a closed Robertson-Walker universe, where the scale factor and the homogeneous mode of the scalar field model the minisuperspace degrees of freedom. They explicitly computed the back-reaction of the inhomogeneous modes on the minisuperspace sector using a coarse-grained effective action and show that the minisuperspace approximation is valid only when this backreaction is small. 

 For   backreaction  due to quantized matter fields, its decoherence effects,  and how the semiclassical limits are reached in quantum cosmology, see,  e.g.,
\cite{Wad86,PadSin90,CalChaosQCDec}.  

Concerning quantum backeaction, specifically the question of  how the inhomogeneous quantum modes backreact on the lower dimensional superspace modes,  and whether the results from mini- or midi-superspace can give a fair representation of the full picture, a recent paper by Bojowald \cite{BojoIR} also expressed such a concern:  ``Even though the BKL scenario allows us to use the classical dynamics of homogeneous models to understand space-time near a spacelike singularity, it is a poor justification of minisuperspace models in quantum cosmology. Using a minisuperspace
model to evolve from a nearly homogeneous geometry at late times to a BKL-like geometry at early times means that we begin with a well-justified, approximate infrared contribution
of the full theory, but then push the infrared scale all the way into the ultraviolet." 

Meanwhile,  we shall wait to see further research results from  the practitioners of loop and spin-foam quantum cosmology  to see how the singularity avoidance in LQG might be altered  by   including the backreaction of inhomgeneous quantum modes. 

In the next section we shall discuss the backreaction of quantum field processes and how that affects the dynamics of  classical background spacetimes.  Before ending,  we mention two backreaction effects, one in inflation, the other in classical GR.   Finelli et al \cite{Finelli}  studied the backreaction of scalar fluctuations on the space-time dynamics in the long wavelength limit using a set of gauge invariant variables.  Below, we add a short interlude  in the classical backreaction research programs for completeness, as it bears on gravitational entropy, albeit less so  on the WCH at the cosmological singularity or bounce. \\

\noindent{\bf Backreaction of inhomogeneous mode in classical GR} 

There are two schools of thought on classical backreaction:  This research  began around 2000, with the work of Buchert \cite{Buchert}   and Ellis with collaborators \cite{Ellis} .  The central themes are summarized in this letter \cite{EllBuc}  where these authors make the observation that "A large-scale smoothed-out model of the universe ignores small-scale inhomogeneities, but the averaged effects of those inhomogeneities may alter both observational and dynamical relations at the larger scale."  It also comments briefly on the relation to gravitational entropy.  A broader representation of their work can be found in these  two reviews \cite{Clarkson,BucRas}  See also  \cite{Sussman} 

Green and Wald, on the other hand, showed  in  a series of papers from 2011-2014 \cite{GreWal} that the  backreaction effects  of classical inhomogenous modes are insignificant.  These issues are related to how one should treat the inhomogeneous modes of  shorter wavelengths, such as using the Brill-Hartle-Isaacson average \cite{BrillHartle,Isaacson,Burnett},    and how much they impact  on  the behavior of the  theory if only the lowest modes are retained.  Buchert, Ellis et al \cite{BucEllIrr} countered  the Green-Wald arguments,  reasserting their claims.     In turn Green and Wald replied to this \cite{GreWalCom} with a  simple derivation of their no backreaction results.



\section{Semiclassical Backreaction supports WCH }

In this last section we examine to what degree is WCH consistent  or contradictory with the results of semiclassical backreaction, namely, the backreaction of quantum field processes involving vacuum polarization or fluctuations on the geometrodynamics near the Planck time,  like those due to the trace anomaly or cosmological  particle creation.  These processes have been studied rather thoroughly in the twenty years 1977-1996 as we mentioned before.  Earlier,  in Sec. 1.2, 2.3 and 3.1, we focused on the avoidance of singularity due to the backreaction effects of the trace anomaly and particle creation.   Here we complete our narrative by adding the considerations of the damping of anisotropy and inhomogeneity.

\subsection{Damping of irregularities ensures a smooth transition, critical for cyclic cosmologies}

Sparing the reader  the details of  the derivations of   regularized  stress energy tensors for particle creation and the calculations of the backreaction effects from the solutions of the semiclassical Einstein equation (which can be found in the papers referred to in earlier sections or in a recent book \cite{HuVer20}), we shall just highlight the key features with relevance to WCH in the following.  More comments about gravitational entropy in this context can be found in \cite{HuPLA83}.

1) Vacuum fluctuations of a quantum field are ubiquitous,  one  need not add them in by hand, nor can one wish them away.  Their effects on the background spacetime have to be included in considerations of whether the universe contracts to a singularity or undergoes a bounce.  

2) Many vigorous quantum field- theoretical  studies of  particle creation in (weakly) anisotropic and inhomogeneous spacetimes show that vacuum particle production is abundant near the Planck time,   their backreaction effects are strong and the dissipation of these irregularities happens swiftly.  (Note the significant qualitative differences between semiclassical backreaction of quantum field processes and  classical backreaction of inhomogeneous modes in  GR mentioned earlier. )

3)   The damping of irregularities  drives down the Weyl curvature with their strong and swift actions.  This seems to suggest that quantum field processes act as an effective protector, maybe even a guarantor,  of WCH.  

4) These quantum field processes act in such a way as to draw  the spacetime into a {\it fixed point} in superspace, a three-geometry of high isotropy and homogeneity,   the special status bestowed by the WCH.    
 
5) The overall  backreaction effects of quantum field processes can be summarized by something resembling a {\it Lenz law}: the production of particles acts in such a way as to diminish their production.  This is because, when the universe is isotropized and homogenized,  it becomes conformally-flat, whence no more conformal particles are produced. (There is ground to believe that in the very  early universe massless particles are overwhelmingly more abundant.) 
 
6) For any aficionado of cyclic cosmology,  in addition to showing that a bounce is possible,  the  `reining-in' of the irregularities near the bounce -- in fact, near every bounce --  is particularly important \cite{BarYam}: 
If the isotropic expansion were unstable as the scale of the universe approaches zero, then huge irregularities and anisotropies could accumulate and the successive cycles would be very different and increasingly anisotropic.  Some powerful and ever-ready damping mechanisms need be there to prevent this from happening.  

In summary,  if we ponder upon the ubiquitous nature of these quantum field processes,  their strongly dissipative effects on the spacetime irregularities,  and the isotropized and homogenized outcome `locked-in' by the Lenz law,  we can see that WCH is more likely to be validated when these factors are included in our considerations.   Add to this the fact that these processes apply both in the expansion and contraction phases of the universe,  rendering both the beginning and ending of a cycle to be smooth,  we can see that quantum field processes are good facilitators of cyclic cosmology.  


\subsection{Quantum Gravity:  Macro spacetime manifold from the interaction of micro constituents}

Going from the Planck scale down toward the singularity or bounce,  even from our rather skimpy description earlier, it may appear to the reader that some  quantum gravity theories, e.g, those based on the WdW equation, Hamiltonian quantum gravity, and loop quantum cosmology are more capable of providing a detailed description of quantum spacetime than others, e.g., strings, spin foams, causal sets.  This is not the case.  One needs to first scrutinize whether some tacit,  commonplace,  yet unproven assumption has crept in, that {\it the macro variables are the same as the micro variables}. One should ask, ``Are they?".   This is a pivotal question,  because by assuming there is no change of variables from micro to macro, one only needs to deal with the quantum to classical transition issue. In my opinion,  a more important and  challenging  question is that of micro to macro transition.  

Theoretically,  this cuts into the  key issue of whether the dynamical variables in GR are `fundamental' all the way from the scale of our universe now, captured amazingly well by  Einstein's theory, to the sub-Planckian scales, and,  for the former group of theories, even all the way down toward the singularity.   What if the  variables are collective variables, that the macroscopic spacetime is emergent, and  that general relativity is an effective theory valid only at large scales and  low energies?  We know Nature has many, possibly an infinite number of,  levels of structure, separated by characteristic scales, and each level can be described by a suitably constructed effective theory.  If so,  the theories we now know of the large scale structures of spacetime will very likely become inapplicable at scales below the Planck length.  

Are we to believe that once a macro-variable is quantized, it becomes a micro-variable?  This is a big leap of faith. In Nature,  a quantized macro-variable  can be  a very different physical entity from a micro-variable, which can be of a quantum nature.  E.g., sound is a collective excitation of atoms. Quantizing them yields  phonons.  But quantization of sound will never lead  to the microvariables we want -- atoms. In fact, these collective variables lose their meanings once the atomic scale is reached.  If  the metric or connection forms are collective variables, quantizing them is the wrong way to discover the sub-Planckian microscopic constituents. In contrast, those theories which propose  certain entities as the microscopic constituents of spacetime, such as strings or causets,  need to show that their  interactions  can indeed produce the macroscopic spacetime we know, thus is the difference between quantum and emergent gravity \cite{E/QG}.

\subsection{Quantum Cosmology: Singularity or bounce  ruled by the interaction of micro-constituents}

Varied as the many schools of thought and practice are, there is agreement that quantum gravity refers  to {\it theories of the microscopic structures of spacetime}. Quantum cosmology is supposed to be the application of  quantum gravity to the description of the universe from the Planck scale down to the singularity or the bounce.  Thus, some  knowledge of the underlying theory of quantum gravity is required.  Many theories of quantum cosmology assume the existence of a metric or connection, and many papers, such as in string cosmology, start  with a statement like,  ``Let us consider strings in the FLRW or de Sitter universe".  Right there, I must say, an ontological issue already shows up.  Where does your spacetime manifold come from?  If we believe that strings or loops are  the fundamental microscopic constituents of spacetime at scales below the Planck length, then, before one talks about string or loop cosmology in terms of the  dynamics of the metric or connection,  one needs to explain how the spacetime manifold, this macroscopic structure of our universe,  befittingly described by general relativity, comes into being.  

It  appears that a clear demonstration of how the spacetime manifold emerges from string  or loop interactions is still  largely lacking.  If one takes the dynamical variables of a macroscopic theory of spacetime and quantizes them, thinking that one would get the microscopic degrees of freedom in this way, while disregarding the possibly huge structural  and  behavioral differences between the micro and the  meso- and macro-scopic emergent structures,  the  claims made pertaining to the nature of a singularity or a bounce in the sub-Planckian length regions may not make much sense or even be outright inconsistent.   

Therefore, unless one can provide a proof that the micro-variables describing sub-Planckian physics are the same as the macro variables of the large scale spacetime we are familiar with,  in our opinion,  all predictions and claims about bounces made in quantum cosmology based on quantizing the macro geometric variables in general relativity need be reassessed.

It is with this broader perspective that we should look at the meaning and  applicability of WCH.   It may very well be that it is enough to examine the  validity and implications of WCH at the Planck scale and not extend it to the sub-Planckian scale, because,  after all,  the Weyl curvature, like all the geometric objects, is well defined only when the  spacetime it lives in has a  manifold structure,  which could be an emergent entity of the underlying QG theories for the microscopic constituents of spacetime. \\

\noindent{\bf Acknowledgment}  I would like to thank Professors Paul Anderson and Paulo Moniz for a careful proof- reading of the completed manuscript with helpful suggestions, and for providing the references for the approach to the Big Crunch and for supersymmetic quantum cosmology, respectively.

\newpage

\end{document}